# Spin wave measurements over the full Brillouin zone of multiferroic BiFeO$_3$

Jaehong Jeong[1], E. A. Goremychkin[2], T. Guidi[2], K. Nakajima[3], Gun Sang Jeon[4], Shin-Ae Kim[5], S. Furukawa[6], Yong Baek Kim[6], Seongsu Lee[5], V. Kiryukhin[7], S-W. Cheong[7], and Je-Geun Park[1,8] #

[1] FPRD Department of Physics & Astronomy, Center for Strongly Correlated Materials Research, Seoul National University, Seoul 151-747, Korea
[2] ISIS Facility, STFC Rutherford Appleton Laboratory, Oxfordshire OX11 0QX, UK
[3] Neutron Science Section, MLF Division, J-PARC Center, Tokai, Ibaraki 319-1106, Japan
[4] Department of Physics, Ewha Womans University, Seoul 120-750, Korea
[5] Neutron Science Division, Korea Atomic Energy Research Institute, Daejeon 305-353, Korea
[6] Department of Physics, University of Toronto, Toronto M5S 1A7, Canada
[7] Rutgers Center for Emergent Materials and Department of Physics and Astronomy, Rutgers University, Piscataway NJ 08854, USA
[8] Center for Korean J-PARC Users, Seoul National University, Seoul 151-747, Korea

Using inelastic neutron scattering technique, we measured the spin wave dispersion over the entire Brillouin zone of room temperature multiferroic BiFeO$_3$ single crystals with magnetic excitations extending to as high as 72.5 meV. The full spin waves can be explained by a simple Heisenberg Hamiltonian with a nearest neighbor exchange interaction (J=4.38 meV), a next nearest neighbor exchange interaction (J'=0.15 meV), and a Dzyaloshinskii-Moriya-like term (D=0.107 meV). This simple Hamiltonian determined, for the first time, for BiFeO$_3$ provides a fundamental ingredient for understanding of the novel magnetic properties of BiFeO$_3$.

Recent discoveries of the so-called multiferroic materials, where the magnetic order and ferroelectric polarization coexist, have led to a surge of interest in this rather unusual class of materials [1]. Intense research activities on these systems are partly motivated by pure intellectual desire to understand a number of the following fundamental questions: why and how the two seemingly disparate ground states can find a particular set of compounds hospitable more than the others. The immense potential for future applications as well as the quest for fundamental principles has made these multiferroic compounds to be one of the most sought-after recent topics in material science [2, 3].

Of all multiferroic compounds, $BiFeO_3$ is arguably one of the most interesting systems with both ferroelectric and magnetic transitions above room temperature: with a Neel temperature at $T_N$=650 K and a ferroelectric transition temperature at $T_C$=1100 K [4, 5], exhibiting one of the largest polarization values, ~ 100 $\mu C/m^2$ [6]. Another interesting point to be made is that when it undergoes an antiferromagnetic ordering at 650 K, an incommensurate structure is formed with an extremely long period of 620 Å [5]. We have recently reported how this incommensurate magnetic structure can be linked to the electric polarisation through a lattice anomaly at $T_N$ [7] with an estimated relative volume change of 0.4% at low temperatures like multiferroic hexagonal manganites [8]. For the benefit of discussion later on, we note that the $FeO_6$ octahedron is distorted in such a way that Fe-O bonds are split into two groups at 5 K: 1.948 and 2.109 Å . Probably because of this splitting, there are two types of super-exchange interactions: one is a nearest neighbour exchange interaction (J) along the Fe-O-Fe bond and the other is a next nearest neighbour exchange interaction (J') along the Fe-O-O-Fe bond.

In the hexagonal notation, the propagation vector of the incommensurate structure is $\mathbf{Q_m}$=[0.0045 0.0045 0] at room temperature with a chiral vector of $\mathbf{e_3}$=[1 -1 0] [5]. Although a pseudo cubic notation has been used in some literatures, here we use the hexagonal notation since it better reveals some symmetry of the spin waves as we discuss later (see Fig. 1a & 1c). Interestingly, this incommensurate phase disappears and becomes a simple G-type antiferromagnetic structure when prepared in thin films, presumably because there is inevitably some residual strain induced due to lattice mismatch with the substrates [9]. There have since been numerous studies reporting various interesting surprises in both bulk and thin film forms of $BiFeO_3$: an unusual photovoltaic effect [10], a light-induced large size-change [11], a strain control of magnetic domain [12], an electric field control of spin wave [13], and a control of magnetism by electric field [14], to name only a few.

Despite the numerous reports of the unusual and highly interesting phenomena found in $BiFeO_3$, understanding of the underlying microscopic spin Hamiltonian is not well established yet. Surely, knowing this microscopic magnetic interaction in its entirety ought to be a firm and proper starting point for deeper and fundamental discussions regarding the magnetic properties. A main reason for this absence of the inelastic neutron scattering data, and hence the lack of the full understanding of the spin waves is associated with several experimental and technical difficulties involved, in particular the fact that it is quite challenging to grow large single crystals of $BiFeO_3$ [10].

In this Letter, we have addressed this fundamental problem of measuring and understanding the full spin waves of $BiFeO_3$ by growing several high quality single crystals, which show a large dielectric polarization **P** of 86 $\mu C/cm^2$ for bulk $BiFeO_3$ with the measured **P** value being close to an expected theoretical value [10, 15]. Although these crystals are quite large with mm-sized dimensions, each of them used alone was not big enough for traditional inelastic neutron scattering experiments. Thus we had to use an assembly of ten single crystals by meticulously co-aligning them all within 3° of one another for all three directions of the hexagonal symmetry with the total mass of 1.9 g at the single crystal neutron diffractometer, FCD of HANARO, Korea. The photo of the assembled samples of ten single crystals is given in Fig. 1b.

Using these samples, we have carried out high resolution inelastic neutron scattering experiments with two time-of-flight spectrometers: one is AMATERAS at J-PARC, Japan and another MERLIN at ISIS, UK. All our experiments were done at 5 K. For technical reasons discussed below, data taken at AMATERAS are useful especially for the low energy part of the spin waves whilst the detailed features of the high energy part of the spin waves are clearer in the data taken at MERLIN. Since we used ten co-aligned crystals with a time-of-flight technique (which has intrinsically relatively poor Q-resolutions), the incommensurate magnetic structure with such a small value of $\mathbf{Q_m}$=[0.0045 0.0045 0] can be safely approximated to a simple G-type antiferromagnet with the following lattice parameters: a=5.573 and c=13.842 Å as shown in Fig. 1a. Therefore we will use this G-type structure for our calculations of the spin waves until we include a Dzyaloshinskii-Moriya-like term.

It is also worth mentioning some technical aspects of the experiments. We carried out the AMATERAS experiment with fixed geometries, i.e. incident neutron beam parallel to the [0 0 1] (c*) or [1 2 0] (b*) axes with the incident energy of 94.154 meV. Thanks to the so-called repetition rate multiplication technique [16], we were also able to have additional data for several other incident energies such as 23.630 and 10.513 meV. In order to determine spin waves from the AMATERAS data, we had to search through the data to determine the energy of the spin waves at Q points concerned, which are covered by the experiment.

In order to overcome this technical difficulty associated with the more traditional method of the time-of-flight technique, we performed the MERLIN experiment while rotating the samples from -90 to +80 degrees with the incident beam parallel to the [0 0 1] direction initially: the scattering plane is on the b*-c* plane. After the experiments, we combined all the 171 individual data files to make a single file containing the full four dimensional information of momentum (Q) and energy (E) of the spin waves. We chose the incident energy of 250 meV for the MERLIN experiment to make the detector coverage of the MERLIN spectrometer wide enough to cover the full Brillouin zone of the anticipated spin waves of $BiFeO_3$, in particular the high energy parts whose detailed features are missing in our AMATERAS data. With this configuration and the sample rotation method, we were able to cover the full Brillouin zone of $BiFeO_3$ and project the spin waves with an energy transfer higher than 30 meV for any directions of the Brillouin zone as we like. As it shall become clearer below, this vast amount of the data in the four-dimensional Q-E space is crucial for us to determine exchange interactions precisely.

Although we can obtain the experimental spin waves for any directions of the Brillouin zone using the MERLIN data, we choose Q points of highest symmetry for presentation here as shown in Fig. 1c. For example, we plot the experimental spin waves for the Γ-M-K-Γ-A directions together with the theoretical results in Fig. 2. The upper contour plot in Fig. 2b is obtained from the combined data of the whole MERLIN experiment while data points denoted by circles are obtained by cutting judiciously the AMATERAS data through each of the related Q-E points. There are two reasons why we do not think that what we measured is not due to phonons. First of all, according to theoretical phonon calculations [17], the experimental dispersion curves are too steep to be acoustic phonon. For example, the acoustic mode only goes up to ~ 10 meV, which is much lower than the top of the spin waves branch at 72.5 meV in our data. Second, the measured excitation becomes weaker when compared at equivalent Q points of different Brillouin zones: with increasing the Q values the intensity of the measured excitations gets reduced following the magnetic form factor.

In order to analyze the data, we started with a minimal Heisenberg Hamiltonian only with nearest neighbor interaction (J). We then calculated our theoretical spin waves using a Holstein- Primakoff transformation. In this model Hamiltonian, the value of J is determined from the total width of the full dispersion. For example, we determined J, which is 4.08 meV, from the energy of the spin waves at the A point (see the insert in Fig. 2b). The dashed line in Figs. 2a & 2b represents the theoretical spin waves calculated using this minimal model Hamiltonian. Surprisingly, this simple Hamiltonian appears to explain most of the essential features observed in the experimental dispersion. However, upon close inspections, this simple model Hamiltonian fails to reproduce some of the detailed features of the experimental data. For example, as shown in Fig. 2b, there are clear discrepancies between the experimental data and the theoretical curves near the M point.

To resolve the discrepancies, we extend the minimal model Hamiltonian by including another exchange interaction term (J') for the next nearest neighbors (n.n.n.) in addition to a Dzyaloshinskii-Moriya-like term in the following Heisenberg Hamiltonian:
$$H = J\sum_{n.n.} \mathbf{S}_i \cdot \mathbf{S}_j + J'\sum_{n.n.n.} \mathbf{S}_i \cdot \mathbf{S}_j - \mathbf{D} \cdot \sum_i (\mathbf{S}_i \times \mathbf{S}_{i+\hat{\delta}}),$$
where the first and second sums run over the nearest neighbors (n.n.) and next nearest neighbors (n.n.n.), respectively for the magnetic unit cell of the G-type structure as shown in Fig. 1a. The third term (**D**) describes a Dzyaloshinskii-Moriya-like interaction with $i + \hat{\delta}$ representing the next nearest neighbor of site *i* along the [1 1 0] axis: **D** is parallel to the chiral vector $\mathbf{e_3}$=[1 -1 0]. We will discuss the effects of the Dzyaloshinskii-Moriya-like term on the spin wave later in the paper. Here we would like to point out one subtle point about our use of the Dzyaloshinskii-Moriya-like term. When we identify the antiferromagnetic vector in the Landau-Ginzburg theory of the local spin order parameter, we can show that the **D** term in our spin Hamiltonian is reduced to the same form of Lifshitz invariant in Ref. 18. So we should note that our **D** term is an *effective* Dzyaloshinskii-Moriya term. See also the magnetic unit cell of a simple G-type structure shown in Fig. 1a. In our calculations, we used a spin parameter S=5/2, an ionic value of $Fe^{3+}$, with quantum corrections.

Using this new Hamiltonian without the Dzyaloshinskii-Moriya-like term, i.e. D=0 meV, we can determine a value of J-2J' by fitting the data at the A point: the energy of the spin waves at the A point is obtained algebraically as $6\tilde{S}(J - 2J')$, where $\tilde{S} = \sqrt{\frac{5}{2}\left(\frac{5}{2} + 1\right)}$. We then varied J and J' values systematically in our calculations to fit the experimental data for all the other directions, while keeping the J-2J' value at 4.08 meV. In particular, the energy at the M and K points can be expressed by the following expressions: $4\tilde{S}\sqrt{2[(J - 2J')^2 - 2J'(J - 2J')]}$ at the M point and $3\tilde{S}\sqrt{3[(J - 2J')^2 - J'^2]}$ at the K point, respectively. From the analysis of the data, we finally obtained the best fitting results with J=4.38 and J'=0.15 meV, respectively. Using this set of parameters, we succeeded in obtaining a better agreement between the experimental data and the theoretical curve than before. For instance, this new theoretical curve (solid line) now explains the small deviations seen in the Q-E regions around the M points in the previous calculations with the nearest neighbor interaction alone (see Fig. 2b). Similar improvements can also be found around many other points of the Q-E space. In Fig. 3, we also display the data together with the theoretical curves for the L-M-L direction, where again we can see the improvement brought about by the inclusion of the additional term J'. In passing, although the inclusion of J' in our calculation improves the fittings, surprisingly the minimal Hamiltonian with a single exchange interaction (J) is a very good approximation to the observed experimental data for most of the Q-E space. The origin of the antiferromagnetic next nearest interaction can be attributed to the large distortion of oxygen octahedron with respect to Fe ions occurring at the ferroelectric transition as we discussed before.

We note that this spin Hamiltonian with two exchange interactions is also consistent with all the available bulk data. For example, our choice of J and J' values produces a theoretical $T_N$=620 K, which is not too far off from the experimental value of 650 K. Moreover, including the Dzyaloshinskii-Moriya-like term with a value of D=0.107 meV produces an angular deviation of 3.24° between neighboring spins along the [1 1 0] direction. This angular deviation corresponds to an incommensurate periodicity of $2\pi\left(\frac{J-4J'}{2D}\right)a \cong 620$ Å, $a$ being the $a$-axis lattice constant. We further performed an ensemble average by the Monte Carlo method by employing the standard Metropolis algorithm on the lattice of $L_{[110]}$=110 and $L_{[1\text{-}10]}$=$L_z$=4 with the periodic boundary conditions. We prepared typical configurations at low temperatures for zero fields by means of a simulated annealing method. We then executed at least $10^5$ Monte Carlo steps per spin and discarded typically $2\times10^4$ steps for equilibration. Near the transition, more care was taken to reach sufficient equilibration and ensemble average by increasing steps to $10^6$. In our Monte-Carlo simulations, the incommensurate phase is found to be stable for the parameters space found by the inelastic neutron scattering.

In our Monte-Carlo calculations, we also simulated various magnetic ground states by varying D values in order to investigate the stability of the incommensurate phase. For this, we included a new term for a single ion anisotropy, $-K\sum_i(S_i^z)^2$. According to our results, the incommensurate ground state is found to be stable over a range of D values as long as K is smaller than a critical value of ~0.009 meV, for the experimentally determined J, J', and D values. When the single ion anisotropy (K) becomes larger than the critical value of 0.009 meV, the simple G-type becomes robust against the incommensurate magnetic structure. Therefore, the disappearance of the incommensurate magnetic phase in thin films can be explained in our model Hamiltonian by the single ion anisotropy value (K) being different from the bulk value via the strain due to lattice mismatches. Since the single ion anisotropy is intrinsically sensitive to local distortions, it is conceivable that Fe in thin films samples under enormous strain might experience different local symmetry from the more ideal broken symmetry of the bulk sample.

Let us now discuss the effects of the Dzyaloshinskii-Moriya-like term (D=0.107 meV) on the spin waves. To study the spin waves in a spiral state, we introduced a rotated coordinate for each spin so that one of the principal axes is aligned with the classical spin direction. We then performed a Holstein-Primakoff transformation for the spin operators in this coordinate. The spin wave dispersion curve is plotted in Fig. 2a together with two other cases discussed before. As one can see in Fig. 2a (see the dotted lines in Fig. 2a), the new theoretical spin waves remain almost unchanged for most of the Q-E space when compared with the spin waves (solid line) calculated for Hamiltonian with nearest (J) and next nearest (J') neighbor interactions. Effects of the Dzyaloshinskii-Moriya-like term can only be visible at the very low energy part of the spin waves near the Γ point. As shown in the inserts of Fig. 2a, the inclusion of the Dzyaloshinskii-Moriya-like term in our calculations splits the low energy modes at the Γ point and opens a gap of about ~1 meV at **Q$_m$**=[0.0045 0.0045 0] for the other two modes. The gapless mode is the phason excitation while the gapped modes arise from the spin-flip processes and thus have in-plane components of the spiral plane defined by **Q$_m$** and the hexagonal c-axis. These modes can be detected by polarized neutron scattering experiments. Unfortunately, since we used time-of-flight techniques with unpolarized neutrons (known to have relatively poor Q resolutions) and with 10 co-aligned samples, the splitting at low energy with such a small **Q$_m$** value cannot be observable within the resolutions of our experiments.

In summary, we have succeeded, for the first time, in measuring the full spin waves of $BiFeO_3$ by co-aligning 10 single crystals, and have determined the two most important exchange parameters, which are the nearest and next-nearest neighbor interactions: J=4.38 and J'=0.15 meV, respectively. Surprisingly, a simple spin Hamiltonian with these two exchange interactions is found to be adequate to explain the measured full spin waves over the entire Brillouin zone. By combining Monte-Carlo calculations, we further estimated an *effective* Dzyaloshinskii-Moriya interaction (D=0.107 meV) responsible for the incommensurate magnetic structure with an extremely long period of 620 Å. According to our spin wave calculations, this Dzyaloshinskii-Moriya-like term affects only the very low energy part of the spin waves.

We thank Jung-Hyun Lee, Hyunok Jung, D. T. Adroja, and C. D. Frost for their contributions at the early stages of this work, and S. Ohira-Kawamura, T. Kikuchi, and Y. Inamura for technical assistance. We also appreciate R. Ewings for helps with HORACE program, and T. G. Perring and M. Arai for helpful discussions. JGP was supported by the National Research Foundation of Korea (Grant Nos. KRF-2008-220-C00012 and R17-2008-033-01000-0) and by the research settlement fund of SNU. GSJ was supported by the NRF grant funded by MEST (Quantum Metamaterials Research Center, R11-2008-053-01002-0). VK and SWC were supported by DOE DE-FG02-07ER46328.

Fig. 1 (Color online) It shows (a) the crystal structure of R3c space group with the magnetic unit cell (dashed line) for a simple G-type structure. The horizontal arrow denoted by **D** indicates the direction of the Dzyaloshinskii-Moriya-like vector: (b) a photo of ten crystals co-aligned within 3° of one another, and (c) Brillouin zone of the hexagonal symmetry with the thick lines indicating some of high symmetry directions, along which the data are presented and discussed in the text.

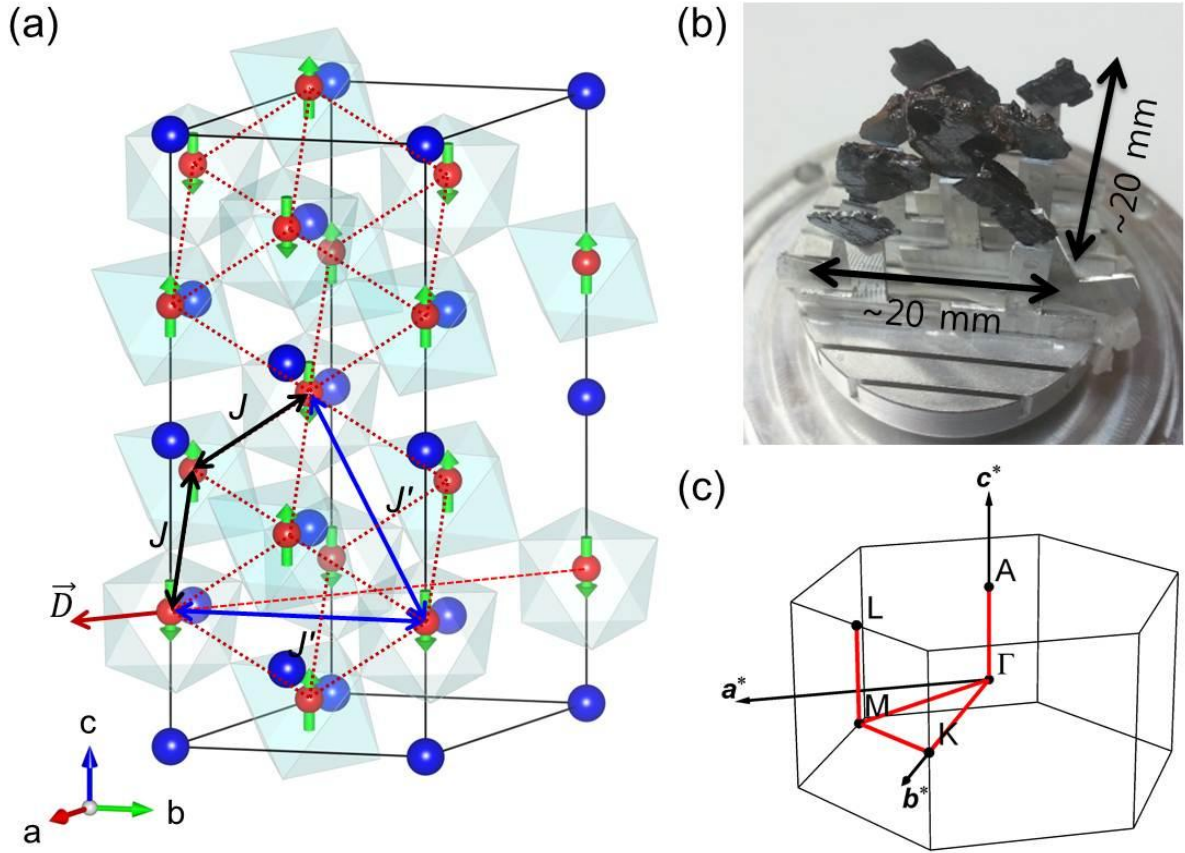

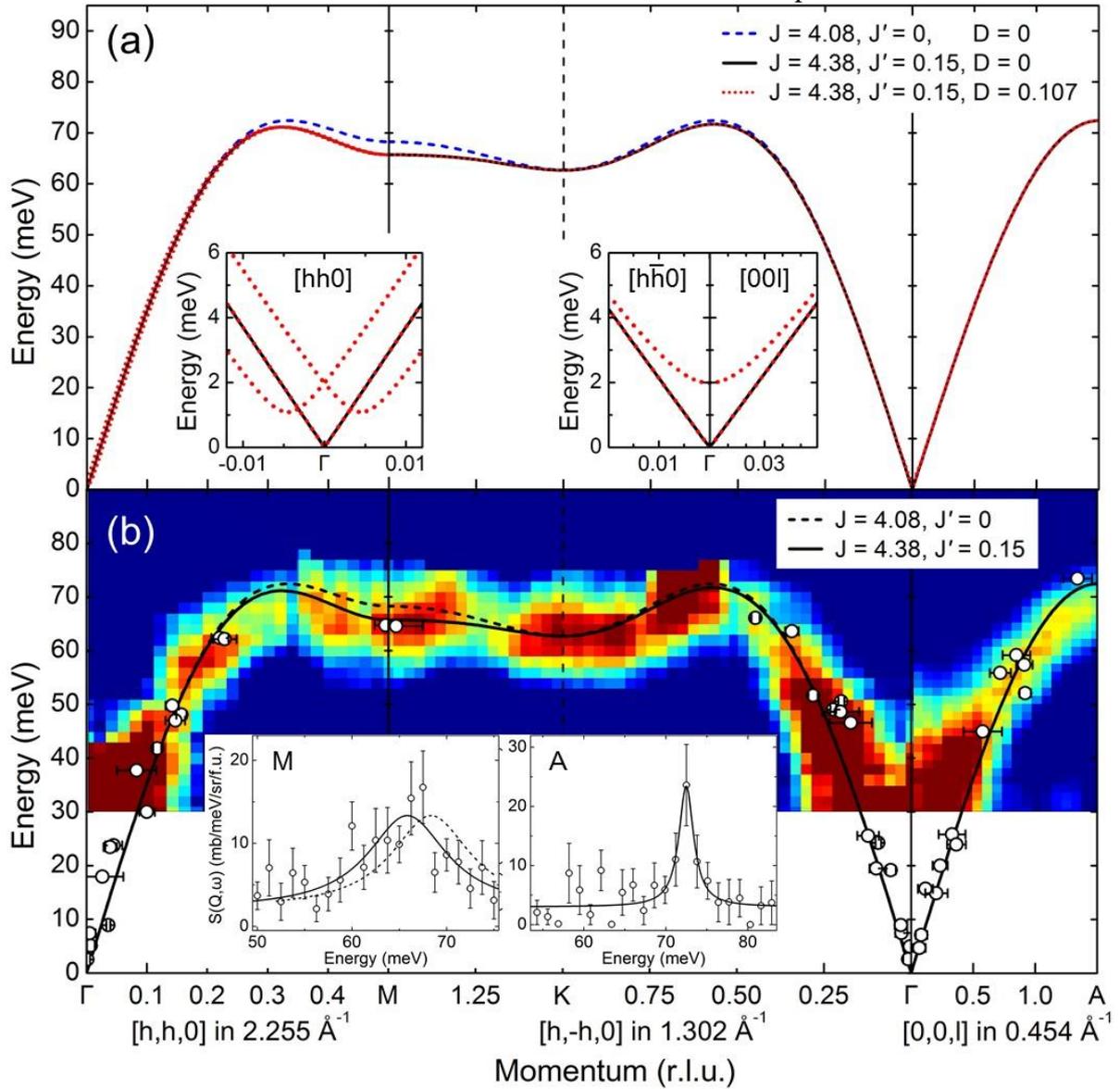

Fig. 2 (Color online) (a) Theoretical spin waves calculated with three different Hamiltonians as discussed in the text. The two inserts are for the blown-up figures of the low energy parts to illustrate the effects of the Dzyaloshinskii-Moriya-like term (D) on the spin waves along the Γ-M and Γ-A directions, respectively. (b) Experimental spin waves measured at AMATERAS beamline (circles) and MERLIN beamline (contour plot) together with the theoretical spin waves (full line) calculated with J=4.38 and J'=0.15 meV: the dashed line is for the theoretical spin waves calculated with Hamiltonian having the nearest neighbor interaction alone. Inserts are for the momentum cut at the M and A points.

Fig 3 (Color online) A contour plot of the spin waves is shown along the [0 0 $l$] direction (L-M-L) with lines for theoretical curves as discussed in the text.

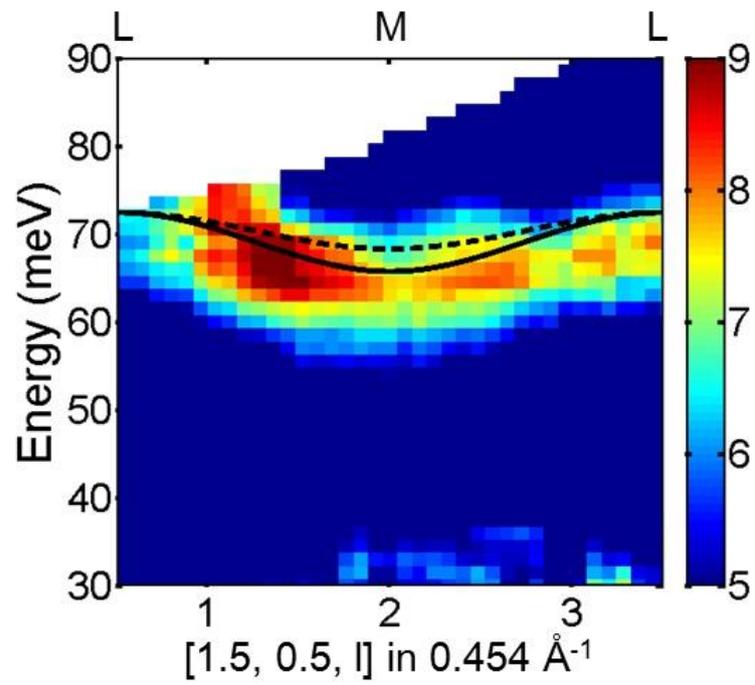